# Band alignment of epitaxial SrTiO$_3$ thin films with (LaAlO$_3$)$_{0.3}$-(Sr$_2$AlTaO$_6$)$_{0.7}$ (001)

Ryan B. Comes[a], Peng Xu[b], Bharat Jalan[b] and Scott A. Chambers[a]

[a]*Physical Sciences Division, Pacific Northwest National Laboratory, Richland, WA, USA*
[b]*Department of Chemical Engineering and Materials Science, University of Minnesota, Minneapolis, MN USA*

**Abstract**

SrTiO$_3$ (STO) epitaxial thin films and heterostructures are of considerable interest due to the wide range of functionalities they exhibit. The alloy perovskite (LaAlO$_3$)$_{0.3}$-(Sr$_2$AlTaO$_6$)$_{0.7}$ (LSAT) is commonly used as a substrate for these material structures due to its structural compatibility with STO and the strain-induced ferroelectric response in STO films grown on LSAT. However, surprisingly little is known about the electronic properties of the STO/LSAT interface despite its potentially important role in affecting the overall electronic structure of system. We examine the band alignment of STO/LSAT heterostructures using x-ray photoelectron spectroscopy for epitaxial STO films deposited using two different molecular beam epitaxy approaches. We find that the valence band offset ranges from +0.2(1) eV to -0.2(1) eV depending on the film surface termination. From these results we extract a conduction band offset from -2.4(1) eV to -2.8(1) eV, indicating that the conduction band edge is more deeply bound in STO and that LSAT will not act as a sink or trap for electrons in the supported film or multilayer.

Epitaxial SrTiO$_3$ (STO) thin films are of considerable interest because of the many intriguing and useful properties they exhibit. STO has been shown to be ferroelectric at low temperatures when grown with compressive in-plane strain on (LaAlO$_3$)$_{0.3}$-(Sr$_2$AlTaO$_6$)$_{0.7}$ (LSAT) (001) substrates and room-temperature ferroelectricity has been demonstrated when STO is grown with in-plane tensile strain on DyScO$_3$.[1] Compressive epitaxial strain in STO films from LSAT substrates has also been shown to reduce the optical band gap of STO.[2] STO also shows extremely high electronic mobilities at low temperature when electron doped with La$^{3+}$ donors on the Sr$^{2+}$ site.[3] Additionally, STO is highly intriguing for its photocatalytic applications due to alignment of the Ti *3d*-derived conduction band with the H$_2$O→H$_2$ half-cell reaction for water splitting.[4] Epitaxial STO films co-doped with Cr$^{3+}$ and La$^{3+}$ grown on LSAT exhibit a reduced optical band gap and enhanced photocatalytic properties,[5] making the surface an ideal candidate for solar photochemistry experiments. Countless other heterostructures and superlattices involving STO have also been grown on LSAT substrates owing to its cubic structure and good lattice match to STO ($a_{LSAT}/a_{STO}$ = 0.991). In many of these systems, a 2-dimensional electron gas (2DEG) is observed within the STO layer,[6,7] though whether any free electrons might diffuse into the LSAT substrate is currently unknown. Collectively, these results underscore the importance of understanding band alignment at this interface lest it become an active electronic component in the overall system.

An understanding of valence and conduction band offsets at semiconductor and insulator interfaces is critically important to the prediction of the behavior of thin films in devices. Much work has been devoted to the examination of STO-Si interfaces and band offsets for transistor gate dielectrics.[8–11] These papers showed that the STO conduction band is nearly perfectly aligned with the Si conduction band in the absence of an interfacial SrO buffer layer, making

STO a very poor gate dielectric on Si due to the likelihood of leakage current across the interface. Similarly, band offset measurements performed using x-ray photoelectron spectroscopy (XPS) on LaAlO$_3$ (LAO)[12] and LaCrO$_3$ (LCO)[13] films grown on STO have helped to elucidate the conductivity (or lack thereof) at these heterojunctions. LSAT films grown on STO have also yielded evidence of 2-dimensional electron gas (2DEG) formation at the interface, attributed to the polar discontinuity between the mixed polarity of LSAT and non-polar STO.[14] In their work, Huang *et al.* postulate a mixture of localized holes and a 2DEG within the STO substrate as a result of the mixed polarity of LSAT, which is claimed to have regions of LaAlO$_3$-like and Sr$_2$AlTaO$_6$-like character. Knowing the actual band offsets for the LSAT-STO heterostructure is thus important for predicting the possibility of carrier transfer into and trapping within LSAT and STO layers of such interfaces.

A series of STO films was grown on LSAT substrates using two different deposition methods, one allowing for *in situ* XPS measurements and the other required a through-air transfer. *In situ* XPS was done using heterojunctions fabricated by conventional oxygen plasma assisted molecular beam epitaxy. In this case, LSAT substrates were sonicated in acetone and isopropanol and cleaned in a UV ozone generator and then immediately loaded into the vacuum chamber. Prior to deposition, the substrates were annealed at 850 °C in a background pressure of $3\times10^{-6}$ Torr activated oxygen (O & O$_2$) generated by an electron cyclotron resonance plasma generator for 15 minutes. The temperature was then lowered to 700 °C and the STO films were deposited using effusion cells and a shuttered approach. A homoepitaxial layer was first grown to calibrate the fluxes to within 1-2% based on a previously published methodology.[15,16] STO films with thicknesses ranging of 5 and 10 unit cells (u.c.) with TiO$_2$ terminations based on the shutter sequence were then deposited, along with a 9.5 u.c. STO film with an SrO termination.

The resulting RHEED patterns and intensity oscillations are shown in Figure 1. The TiO$_2$-terminated surface shows half-order diffraction peaks consistent with a 2×1 surface reconstruction along the [100] azimuth, which is consistent with our previous observations and others.[17–19] We regularly observe these features in the shuttered growth of STO films and have correlated their presence with a TiO$_2$ termination via angle-resolved XPS (ARXPS) measurements of homoepitaxial STO films. No reconstruction features are present for the SrO-terminated film, but we have verified that the surface is primarily SrO terminated via (ARXPS). Immediately following film growth, the samples were transferred under ultra-high vacuum to an appended XPS chamber for band alignment measurements. In the second approach, 5 and 10 u.c. samples were prepared via a hybrid molecular beam epitaxy approach which uses a titanium isopropoxide chemical precursor for Ti, an elemental source to supply Sr and a RF plasma source to supply oxygen.[20] This growth approach is known to result in a MBE growth window, i.e. for a range of Sr:Ti beam flux, only stoichiometric SrTiO$_3$ grows.[21] The hybrid MBE grown samples were transferred through air and measured using the same XPS system, both as received and after room-temperature exposure to activated oxygen in the appended conventional MBE system.

XPS was carried out using a monochromatic Al Kα x-ray source with a VG Scienta R3000 electron energy analyzer. Measurements were made at normal emission and 70° off-normal for additional surface sensitivity. LSAT substrates cleaned by using the same 850 °C heat treatment in activated oxygen were used to measure the binding energies of the Ta 4$d$ core levels and the valence band, in order to determine the energy difference between the Ta 4$d_{5/2}$ peak and the valence band maximum (VBM) for bulk LSAT(001). These spectra are shown in Figure 2 and yield a $(E_{Ta4d} - E_V)_{LSAT}$ value of 227.62(5) eV. The energy difference between the Ti 2$p_{3/2}$ and Ta 4$d_{5/2}$ peaks for each heterojunction, $(E_{Ti2p} - E_{VTa4d})_{HJ}$, was then measured and combined with the energy separation between the Ti $2p_{3/2}$ peak and the VBM for bulk STO(001), $(E_{Ti2p} - E_V)_{STO}$, which equals 455.95(4) eV[12], in order to determine the valence band offset (VBO). The LSAT optical gap of 5.8 eV was determined via spectroscopic ellipsometry measurements performed on a reference LSAT substrate and was used to estimate the conduction band offset from the VBO and the band gap of STO, 3.2 eV.

A summary of the results is found in Table 1. We find a small VBO of -0.2(1) eV for 5 and 10 u.c. $TiO_2$-terminated films, whereas the SrO-terminated and hybrid MBE grown samples were found to exhibit small but slightly positive VBOs of +0.1(1) to +0.2(1) eV. A (positive) negative VBO indicates that the STO VBM lies at (higher) lower energy than that of LSAT, as seen in the insets for Fig. 3. Using the known band gaps for STO and LSAT, we use these VBOs to estimate conduction band offsets (CBO) ranging from -2.4(1) to -2.8(1) eV. The XPS data and a schematic of the result for the 10 u.c. $TiO_2$-terminated and 9.5 u.c. SrO-terminated samples are shown in Figure 3.

The large conduction band offset between STO and LSAT clearly demonstrates that for any STO-based heterostructures grown on LSAT, electrons will not diffuse into the LSAT. However, for p-type heterostructures, holes would easily diffuse across the interface and could become trapped in the LSAT. These results can be intuitively understood given that both the STO and LSAT valence bands have formally $d^0$ cations ($Ta^{5+}$, $Al^{3+}$ and $Ti^{4+}$), so that the valence bands are almost exclusively of O *2p* character and any holes induced on the oxygen ions would readily migrate across the interface.

The small change in the VBO upon change in surface termination is a curious result worthy of further consideration. If the films are slightly off-stoichiometry, it is possible that excess Sr or Ti is present at the surface, resulting in localized patches of SrO or $TiO_2$. This type of cation surface rearrangement to accommodate off-stoichiometry has been observed in SrTiO$_3$ previously through synchrotron x-ray diffraction[22] and in LaAlO$_3$ using RHEED to observe different surface reconstructions.[23] Previous theoretical papers have suggested that the VBO between SrO and STO is 0.2 eV,[10] which is on the scale of our observed difference. However, it is not clear how a surface band offset between the excess SrO and the STO film surface would affect the buried STO-LSAT interface. We do not observe any variation in Ti *2p* line shape or peak width that might reveal small amounts of band bending at the surface due to the differing surface termination. Such band-bending could confuse the results due to the far greater sensitivity of XPS to the film surface than to the buried interface, but we see no evidence to suggest that it occurs.

Given that surface effects are unlikely to be the direct cause of the variation in measured band offset, it is more likely that variability in the properties of the LSAT surface are the cause, and that the connection to surface termination is coincidental. The LSAT(001) surface can exhibit a wide variety of different terminations (A site and B site terminations with all four cations present in different proportions) and surface defects, such as SrO island formation after air annealing.[24,25] While SrO islands would be readily apparent through RHEED pattern analysis, precise determination of the surface termination is more challenging. Angle-resolved XPS measurements for an LSAT substrate annealed *in situ* prior to growth showed a mixed termination, with no preference for A site or B site termination, and no surface segregation of particular cations. Post-growth XPS showed a primarily SrO-terminated film in the case of the 9.5 u.c. film and a mixed termination for the 10 u.c. film, consistent with the initial mixed termination of the substrate and a stoichiometric film. The LSAT(001) surface has both positive and negative polarity due to regions behaving like LAO and regions behaving like Sr$_2$AlTaO$_6$.[14]

Differing terminations of a polar/non-polar interface such as the one present here has been shown to cause variations in the degree of intermixing in LaAlO$_3$-STO heterostructures.[26] We suggest that variability in surface termination and stoichiometry has a minor effect on the VBO, either directly or via subtle differences in cation mixing at the interface. Further analysis with differing LSAT surface terminations could help to clarify these observations.

To summarize, we have determined the valence and conduction band offsets between SrTiO$_3$ epitaxial thin films and (LaAlO$_3$)$_{0.3}$-(Sr$_2$AlTaO$_6$)$_{0.7}$ substrates using x-ray photoelectron spectroscopy and spectroscopic ellipsometry. We find that the VBO is approximately 0 eV, but that the surface termination can produce variations of ~±0.2 eV. Based on the measured optical band gap of LSAT (5.8 eV), we infer a conduction band offset of between 2.4 and 2.8 eV. This result indicates that any itinerant electrons in the system will be confined to the film structure and that any electronic contributions from the substrate can be safely discounted. However, the variation in the valence band offset with surface termination indicates that the termination and film stoichiometry may affect interfacial intermixing and the resulting valence band alignment. Care should thus be taken to justify any claims regarding the interface between these two materials that assume an ideal interface.

**Acknowledgements**

R.B.C. was supported by the Linus Pauling Distinguished Post-doctoral Fellowship at Pacific Northwest National Laboratory (PNNL LDRD PN13100/2581). S.A.C was supported at PNNL by the U.S. Department of Energy, Office of Science, Division of Materials Sciences and Engineering under Award #10122. The PNNL work was performed in the Environmental Molecular Sciences Laboratory (EMSL), a national science user facility sponsored by the Department of Energy's Office of Biological and Environmental Research and located at Pacific

Northwest National Laboratory. The work at the University of Minnesota was supported primarily by the MRSEC Program of the National Science Foundation under award number DMR-1420013 and in part by NSF DMR-1410888. We also acknowledge use of facilities at the NSF-funded UMN Characterization Facility and the Nanofabrication Center.

Table 1. Core-level binding energies and band offsets for STO/LSAT heterojunctions

| STO layer thickness | Ti $2p_{3/2}$ (eV) | Ta $4d_{5/2}$ (eV) | $\Delta E_V$ (eV) † | $\Delta E_C$ (eV) § |
|---|---|---|---|---|
| 5 u.c.* | 456.30(2) | 227.8(1) | -0.2(1) | -2.8(1) |
| 9.5 u.c.* | 455.91(2) | 227.7(1) | +0.1(1) | -2.5(1) |
| 10 u.c.* | 455.97(2) | 227.4(1) | -0.2(1) | -2.8(1) |
| 5 u.c.** | 456.24(2) | 228.0(1) | +0.1(1) | -2.5(1) |
| 10 u.c.** | 455.77(2) | 227.6(1) | +0.2(1) | -2.4(1) |
| 10 u.c.*** | 455.66(2) | 227.7(1) | +0.4(1) | -2.2(1) |

\* Conventional shuttered MBE growth, measured *in situ*.
\*\* Hybrid MBE growth, measured *ex situ*.
\*\*\* Hybrid MBE growth, measured *ex situ* after cleaning in oxygen plasma at room temperature.
† $\Delta E_V = (E_{Ti2p} - E_V)_{STO} - (E_{Ta4d} - E_V)_{LSAT} - (E_{Ti2p} - E_{Ta4d})_{HJ}$
§ $\Delta E_C = \Delta E_g - \Delta E_V$ where $\Delta E_g = E_g^{STO} - E_g^{LSAT}$ and the STO and LSAT gaps are taken to be 3.2 and 5.8 eV, respectively.

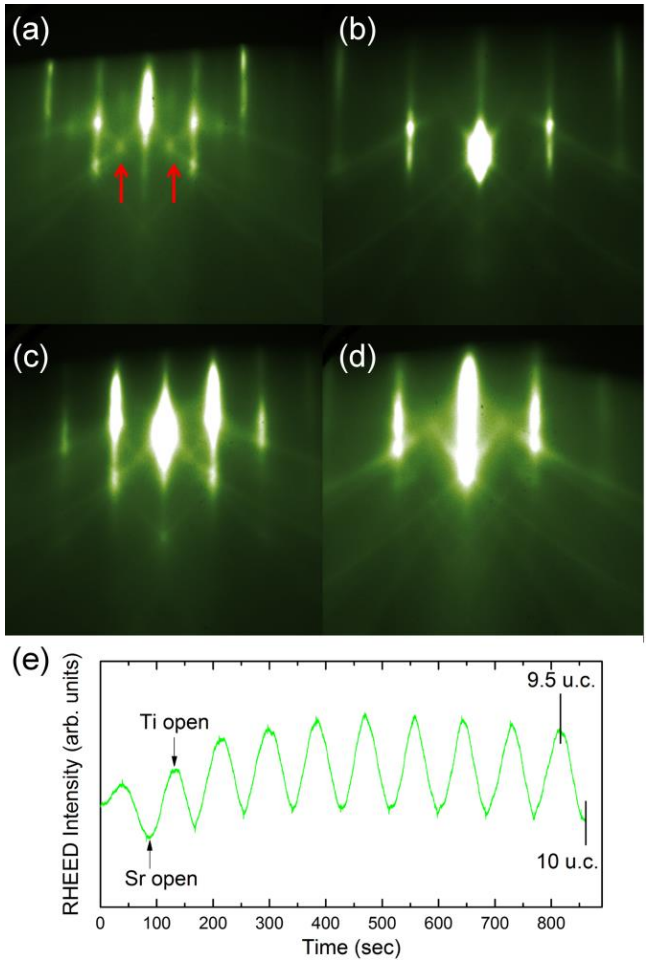

**Figure 1.** RHEED patterns of SrTiO$_3$ films grown on LSAT for: (a) 10 u.c. TiO$_2$-terminated STO along [100] azimuth and (b) along [110] azimuth; (c) 9.5 u.c. SrO-terminated STO along (10) azimuth and (d) along (11) azimuth. (e) RHEED intensity oscillations for 10 u.c. film with notes showing the shuttering process for both films. Arrows point to reconstruction features commonly associated with TiO$_2$ terminations for the respective films.

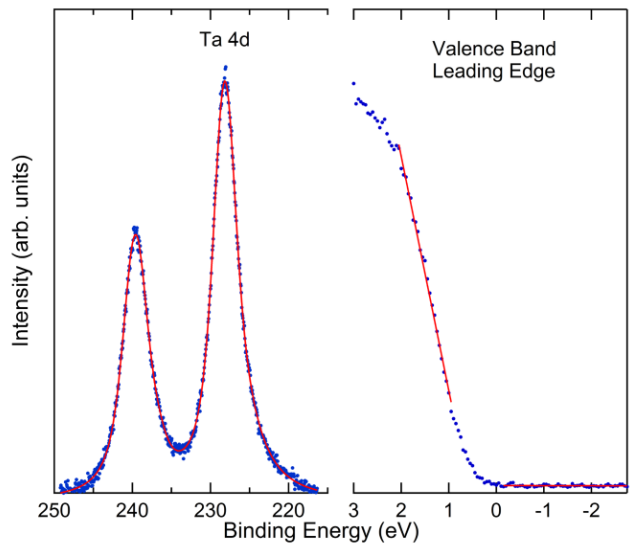

**Figure 2.** Ta 4d core-level and valence-band spectra for LSAT substrate, with fits to the data used to calculate the energy difference between the valence band maximum relative to the Ta $4d_{5/2}$ peak.

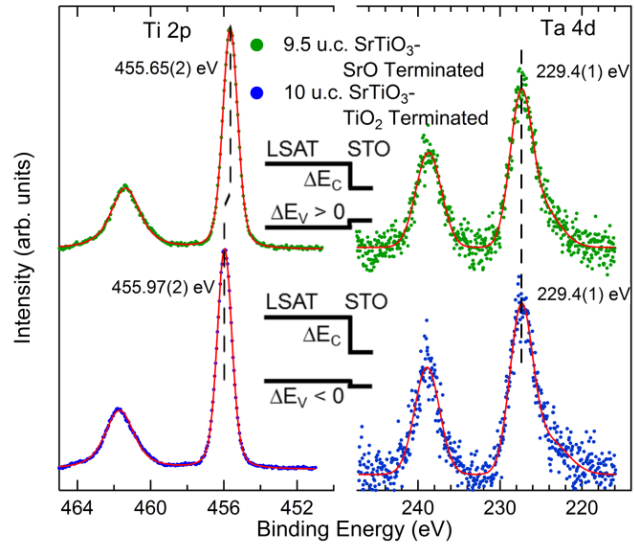

**Figure 3.** Ti 2p and Ta 4d core level spectra for 10 u.c. $TiO_2$-terminated and 9.5 u.c. SrO-terminated films with fits to the data used to determine valence band offsets.